\documentclass{IEEEtran}

\usepackage{amsbsy}
\usepackage{amsmath}
\usepackage{amssymb}
\usepackage{bm}
\usepackage{cite}
\usepackage{graphicx}
\usepackage{mathbbol}
\usepackage{theorem}
\usepackage{tikz}
\usetikzlibrary{calc}
\usetikzlibrary{arrows}

\newcommand{\insertfig}[2]{
\begin{figure}\centering
\includegraphics[width=.9\columnwidth]{#1.pdf}
\caption{#2}\label{#1.fig}\end{figure}}


\newtheorem{lemma}{Lemma}[section]

\newtheorem{theorem}{Theorem}
\theorembodyfont{\rmfamily}

\newtheorem{remark}{Remark}[section]
\def\ben{\begin{enumerate}}
\def\beq{\begin{equation}}
\def\beqa{\begin{eqnarray}}
\def\bit{\begin{itemize}}
\def\een{\end{enumerate}}
\def\eeq{\end{equation}}
\def\eeqa{\end{eqnarray}}
\def\eit{\end{itemize}}
\def\dst{\displaystyle}
\def\non{\nonumber\\}

\DeclareMathAlphabet{\mathsfbf}{OT1}{cmss}{sbc}{n}
\def\A{\mathcal{A}}
\def\Am{\bm{A}}
\def\Amc{\check{\Am}}

\def\Bm{\bm{B}}
\def\Bmc{\check{\Bm}}

\def\betav{\bm{\beta}}

\def\Cm{\bm{C}}

\def\CN{\mathcal{CN}}

\def\diag{{\rm{diag}}}

\def\H{\mathsfbf{H}}

\def\Hm{{\bm{H}}}

\def\Id{\bm{I}}

\def\Km{\bm{K}}

\def\L{\mathcal{L}}

\def\Lam{\bm{\Lambda}}

\def\Lamc{\check{\Lam}}

\def\Lamt{\widetilde{\Lam}}

\def\lamv{\bm{\lambda}}

\def\lamvt{\tilde{\lamv}}
\def\muv{\bm{\mu}}

\def\Pim{\bm{\Pi}}

\def\Qm{\bm{Q}}
\def\Qmc{\check{\Qm}}

\def\Qmt{\widetilde{\Qm}}

\def\rank{\mathrm{rank}}

\def\Sgm{\bm{\Sigma}}

\def\T{\mathsfbf{T}}

\def\tr{\mathop{\mathrm{tr}}}

\def\Um{\bm{U}}
\def\Umc{\check{\Um}}
\def\Umt{\widetilde{\Um}}

\def\Xm{\bm{X}}
\def\Xmc{\check{\Xm}}

\def\xv{\bm{x}}

\def\yv{\bm{y}}
\def\yvt{\tilde{\yv}}

\def\0{\bm{0}}
\def\1{\bm{1}}

\def\zv{\bm{z}}

\def\zvt{\tilde{\zv}}

\begin{document}

\title{Information Rate Optimization for Non-Regenerative MIMO Relay Networks with a Direct Link$^\sharp$}
\author{Giorgio Taricco}

\maketitle
\def\thefootnote{$\sharp~~$}
\footnotetext{
Giorgio Taricco (gtaricco@ieee.org) is with Politecnico di Torino (DET).
A limited part of these results have been presented in IEEE ISIT 2021 \cite{isit21}.
}
\def\thefootnote{\arabic{footnote}}

\begin{abstract}
We consider the optimization of a two-hop relay network based on an amplify-and-forward Multiple-Input Multiple-Output (MIMO) relay.
The relay is assumed to derive the output signal by a Relay Transform Matrix (RTM) applied to the input signal.
Assuming perfect channel state information about the network at the relay, the RTM is optimized according to two different criteria:
{\bf\em i)} network capacity;
{\bf\em ii)} network capacity based on Orthogonal Space--Time Block Codes.
The two assumptions have been addressed in part in the literature.
The optimization problem is reduced to a manageable convex form, whose KKT equations are explicitly solved.
Then, a parametric solution is given, which yields the power constraint and the capacity achieved with uncorrelated transmitted data as functions of a positive indeterminate.
The solution for a given average power constraint at the relay is amenable to a \emph{water-filling-like} algorithm, and extends earlier literature results addressing the case without the direct link.
Simulation results are reported concerning a Rayleigh relay network and the role of the direct link SNR is precisely assessed.
\end{abstract}

\section{Introduction}

Wireless communication systems have been using relaying techniques for several decades in order to extend the range coverage of radio networks.
Among the benefits, relays help to combat shadowing and fading effects which may limit the signal propagation in wireless environments.
Back in the day, the concept of relaying has been rationalized, from an information theoretical point of view, by the introduction of the three-terminal channel model (source-relay-destination) in the seminal papers by Van Der Meulen~\cite{vdm71} and Cover and El Gamal~\cite{cover79}, which determined the achievable rate under several operating conditions.
During the last two decades, several results emerged in the framework of single and multiple antenna systems.
As far as single-antenna systems are concerned,
Sendonaris \emph{et al.} studied the effects of relaying as a user cooperation diversity technique to increase the cellular coverage of third-generation systems based on CDMA \cite{send03};
Nabar \emph{et al.} investigated different time-division multiple-access-based cooperative protocols with relay terminals operating in either amplify-and-forward or decode-and-forward modes by using the achievable rate as the metric of interest \cite{nab04};
Laneman \emph{et al.} proposed low-complexity cooperative diversity protocols to combat multipath fading in wireless networks by using
several strategies including fixed relaying schemes [Amplify-and-Forward (AF) and Decode-and-Forward (DF)], adaptive relaying schemes, and incremental relaying schemes based upon limited feedback from the destination terminals \cite{lane04};
Host-Madsen \emph{et al.} provided lower and upper bounds to the outage and ergodic capacity of a three-terminal wireless relay channel in Rayleigh fading while taking into account practical constraints at the relay node and the impact of power allocation \cite{host05}.

Relaying based on Multiple-Input Multiple-Output (MIMO) wireless terminals has been studied in~\cite{wang05,tang07}.
Specifically, joint transmission and reception at the relay was addressed in the paper by Wang \emph{et al.}~\cite{wang05} but, as pointed out by Tang and Hua, \cite{tang07}, it may entail unwanted side effects since, typically, the transmitted signal power at the relay overshadows the power of the received signal.
As a result, a more practical approach consists of keeping the reception and transmission processes at the relay orthogonal with respect to each other.
Orthogonality can be implemented by operating the system in a two-hop time division or by frequency-domain division multiple access scheme.

Focusing on the relayed signal, two basic approaches have been considered in the literature, which can be classified as \emph{regenerative} or \emph{non-regenerative}.
The regenerative approach consists of rebuilding the transmitted signal after decoding the received signal, and is commonly referred to as \emph{decode-and-forward} (DF).
The non-regenerative consists of forwarding the received signal after amplification, thereby including the received noise.
This latter approach is commonly referred to as \emph{amplify-and-forward} (AF).
For single-antenna systems, it has been observed that AF schemes are advantageous in terms of achievable diversity order with respect to DF schemes while the situation is not clearly understood as far as concerns capacity.
Nevertheless, AF schemes offer a number of benefits making them preferable to DF schemes~\cite{tang07}.
More recently, it has been pointed out that AF schemes enable to retain the soft information of the transmitted signal and guarantee a limited signal delay at the same time~\cite{gaz14,lee14,liang17}.

In the framework of MIMO-AF relay schemes, the transmitted signal is obtained by the joint amplification of the different received signal components so that it can be characterized by a \emph{Relay Transform Matrix} (RTM), which derives the transmitted signal vector through multiplication by the received signal vector.
Following the classification introduced by Tang and Hua~\cite{tang07}, we consider three operating schemes for the MIMO relay system considered:
$i)$ Direct Link without Relay;
$ii)$ Relay without Direct Link; and
$iii)$ Relay with Direct Link.
A key contribution from~\cite{tang07} is the derivation of the information-theoretically optimum RTM (i.e., maximizing the system capacity) for the second operating scheme (Relay without Direct Link).
The authors also considered the more general third operating scheme (Relay with Direct Link) but didn't find the optimum RTM in this case and claimed this case to be an open problem~\cite[p.1400]{tang07}.
This operating scheme was also considered by Shariat and Gazor~\cite{gaz14}, who focused on the optimization of the capacity constrained to the use of Orthogonal Space--Time Block Codes (OSTBC).
Their approach is also equivalent to the maximization of the overall Signal-to-Noise Ratio (SNR).

Several works considered the Relay with Direct Link scheme (also in the presence of a precoder at the source) with the goal of obtaining lower or/and upper bounds to the achievable rate \cite{palla1,palla2,palla3,palla4}.
More recently, some work considered the impact of imperfect channel state information, known according to its statistic distribution, and obtained upper bounds to the ergodic capacity \cite{palla5}.

The RTM optimization has also been studied under a different optimization criterion, namely, the Minimum Mean-Square Error (MMSE) minimization.
Several works adopted this approach, such as \cite{song12,kong14,he16}.
In general, MMSE optimization is not equivalent to capacity optimization but this approach lends itself to a simpler solution of the optimization problem.

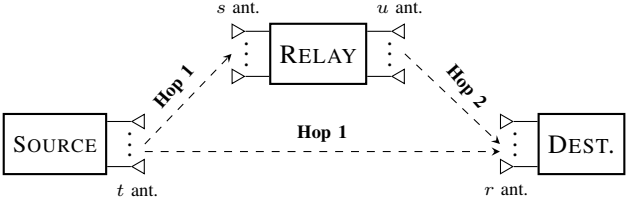
\begin{figure}
\begin{center}
\begin{tikzpicture}[scale=1,>=stealth,
blk/.style={fill=white,minimum width=1cm,minimum height=.8cm,draw,thick}
]
\path(0,0)node[blk](s){\small\sc Source}(3.5,1.2)node[blk](r){\sc Relay}(7,0)node[blk](d){\sc Dest.};
\def\rant#1{
\draw[>=open triangle 60,-.<]($(#1.east)+(0,.3)$)--+(.5,0);
\node at($(#1.east)+(.3,.1)$){$\vdots$};
\draw[>=open triangle 60,-.<]($(#1.east)+(0,-.3)$)--+(.5,0);}
\def\lant#1{
\draw[>=open triangle 60,-.<]($(#1.west)+(0,.3)$)--+(-.5,0);
\node at($(#1.west)+(-.3,.1)$){$\vdots$};
\draw[>=open triangle 60,-.<]($(#1.west)+(0,-.3)$)--+(-.5,0);}
\rant{s}
\rant{r}
\lant{r}
\lant{d}
\draw[dashed,->]($(s.east)+(.5,0)$)--($(r.west)+(-.5,0)$)node[midway,above,rotate=49]{\scriptsize\bfseries Hop 1};
\draw[dashed,->]($(s.east)+(.5,-.1)$)--($(d.west)+(-.5,-.1)$)node[midway,above,rotate=0]{\scriptsize\bfseries Hop 1};
\draw[dashed,->]($(r.east)+(.5,0)$)--($(d.west)+(-.5,0)$)node[midway,above,rotate=-49]{\scriptsize\bfseries Hop 2};
\node[below right]at(s.south east){\scriptsize $t$ ant.};
\node[above left]at(r.north west){\scriptsize $s$ ant.};
\node[above right]at(r.north east){\scriptsize $u$ ant.};
\node[below left]at(d.south west){\scriptsize $r$ ant.};
\end{tikzpicture}
\end{center}
\caption{System block diagram.
Transmission occurs in two time/frequency slots (hops) so that the received signal at the destination arrives alternately from the source and from the relay.}
\label{sys.fig}
\end{figure}

In this work we present an algorithm to derive the RTM optimizing the capacity of a two-hop relay network.
The case considered here is more general than~\cite{tang07,gaz14} for many reasons.
First of all, we allow the number of transmit and receive antennas at the relay to be arbitrarily different, as well as the channel matrix ranks.
On the contrary, it was assumed in \cite{tang07} that the number of transmit and receive antennas at the relay was the same and \cite{gaz14} assumed that the rank of the source to destination channel matrix ($\Hm_1$ in this paper) was equal to the number of receive antennas, so that the case $t<r$ (see Fig.\ \ref{sys.fig} for the definitions) was not included.
Additionally, the solution presented here applies to the joint direct link and relay transmission case (labeled as ``Case (C) Relay With Direct Link'' in \cite{tang07}), recognized as an \emph{open problem} by the authors of \cite{tang07}.
Our solution can be obtained, for a specific relay power constraint, by resorting to a water-filling-like algorithm, bearing some similarity with~\cite[Sec.IV]{tang07} (which is nevertheless not applicable to this case).
For validation purposes, we report numerical simulation results coherent with those from~\cite[(B) Relay Without Direct Link]{tang07} by forcing the direct link channel matrix to zero.
Then, we extend the analysis by considering also the case of a full relay network, including the direct link, first with an overall constant number of antennas in the relay network, next, with different number of antennas.
A \emph{distinguishing feature} of this work with respect to earlier literature results (\emph{e.g.}, \cite{palla1,palla2,palla3,palla4,palla5}) is the consideration of the actual capacity instead of some lower or upper bound (which nevertheless provide valuable contributions).

Summarizing, the paper organization is as follow.
Section~\ref{sysmod.sec} introduces the system model for the MIMO relay network with all relevant parameters which characterize it completely.
Then, Section \ref{RTM.sec} solves the optimization problems corresponding to capacity and OSTBC-capacity maximization in the fully general case of arbitrary channel matrix ranks and dimensions.
Section \ref{RTM.Th1.sec} addresses capacity optimization and extends the work of \cite{tang07}.
Section \ref{param.sec} proposes the relevant parametric solution.
Section \ref{RTM.Th2.sec} addresses OSTBC-capacity optimization and extends partly the work of \cite{gaz14} and
Section~\ref{param2.sec} proposes the relevant parametric solution.
Section \ref{results.sec} collects three types of relay network scenarios to illustrate the application of the theoretical results of the previous section.
The first scenario consists of a relay network without the direct link and is considered for validation and comparison with the results of \cite{tang07}.
The other scenarios consider a full relay network with constant number of antennas (where \cite{gaz14} is applicable as far as OSTBC-capacity is concerned) an a second scenario with different numbers of antennas (where \cite{gaz14} is not applicable even in the case of OSTBC-capacity).
Concluding remarks are collected in Section \ref{conclusions.sec}.

\section{System Model}\label{sysmod.sec}

We consider a MIMO relay network consisting of three nodes:
the source (S) equipped with $t$ transmit antennas;
the destination (D), equipped with $r$ receive antenna;
and the relay (R), equipped with $u$ transmit and $s$ receive antennas.
The channel matrices corresponding to the three different links of interest are labeled as $\Hm_0$ (S$\to$D), $\Hm_1$ (S$\to$R), and $\Hm_2$ (R$\to$D).
The system operates in \emph{two-hop relaying} mode: the source transmits during the first hop and the relay during the second hop.
The average power transmitted by the source and the relay are upper bounded by $P_1$ and $P_2$, respectively.
We assume that the relay applies a $u\times s$ \emph{Relay Transformation Matrix} (RTM) $\Xm$ to the received signal before forwarding it to the destination in the second hop.
The resulting channel equations are given as follows:
\begin{align}\label{ch.eq}
\begin{array}{r@{\,}r@{\,}r@{\,}ll}
\yv_0&=&\Hm_0\xv&+\zv_0&\text{(Hop 1, S$\to$D)}\\
\yv_1&=&\Hm_1\xv&+\zv_1&\text{(Hop 1, S$\to$R)}\\
\yv_2&=&\Hm_2\Xm\yv_1&+\zv_2&\text{(Hop 2, R$\to$D)}\\
&=&\Hm_2\Xm\Hm_1\xv&+\Hm_2\Xm\zv_1+\zv_2
\end{array}
\end{align}
We assume, w.l.o.g., that the received noise components are iid (otherwise, we can pre-multiply the received vectors and channel matrices by the inverse matrix square roots of the corresponding noise correlation matrices).
Then,%
\footnote{
The notation $\zv\sim\CN(\muv,\Sgm)$ is associated to the circularly-symmetric complex Gaussian distribution of the random vector $\zv$ and the corresponding pdf is defined by
$f_{\zv}(\zv)=\det(\pi\Sgm)^{-1}\exp[-(\zv-\muv)^\H\Sgm^{-1}(\zv-\muv)]$.
}
\begin{equation}
\zv_0,\zv_2\sim\CN(\0,\Id_r),\quad\zv_1\sim\CN(\0,\Id_s)
\end{equation}
The equivalent channel equation becomes
\begin{equation}
\yv=\begin{pmatrix}\Hm_0\\\Hm_2\Xm\Hm_1\end{pmatrix}\xv
+\begin{pmatrix}\zv_0\\\Hm_2\Xm\zv_1+\zv_2\end{pmatrix}
\end{equation}
After decorrelating the second hop noise component, the channel equation can be written as follows:
\begin{equation}
\yvt=\begin{pmatrix}\Hm_0\\(\Hm_2\Xm\Xm^\H\Hm_2^\H+\Id_r)^{-1/2}\Hm_2\Xm\Hm_1\end{pmatrix}\xv+\zvt
\end{equation}
where $\zvt\sim\CN(\0,\Id_{2r})$.
This channel equation leads directly to the equation representing the capacity of the relay network reported in the following eq.\ \eqref{C.eq}.

\section{RTM Optimization}\label{RTM.sec}

In this section we address the calculation of the optimum RTM based on the assumption that the relay \emph{knows} all the channel matrices involved in eq.\ \eqref{ch.eq}.
Specifically, we look for the RTM which maximizes the two-hop relay channel capacity.

\subsection{Optimum RTM}\label{RTM.Th1.sec}

In the absence of Channel State Information at the Transmitter (CSIT), the capacity is achieved when $\xv\sim\CN(\0,\frac{P_1}{t}\Id_t)$ and is given by
\begin{align}\label{C.eq}
C=\log_2
&\det\bigg\{\Id_{t}+\frac{P_1}{t}\Big[\Hm_0^\H\Hm_0+\Hm_1^\H\Xm^\H\Hm_2^\H\non
&(\Id_r+\Hm_2\Xm\Xm^\H\Hm_2^\H)^{-1}\Hm_2\Xm\Hm_1\Big]\bigg\}
\end{align}

The average power constraint at the relay can be expressed in terms of the RTM $\Xm$ and the channel matrices as follows:
\begin{align}\label{constr.eq}
\tr\bigg\{\Xm\Xm^\H&+\frac{P_1}{t}\Xm\Hm_1\Hm_1^\H\Xm^\H\bigg\}\non
&=\tr\bigg\{\Xm\bigg(\Id_s+\frac{P_1}{t}\Hm_1\Hm_1^\H\bigg)\Xm^\H\bigg\}
\le P_2.
\end{align}

The optimum RTM (maximizing the capacity \eqref{C.eq} under the constraint \eqref{constr.eq}) is given by the following Theorem.
\begin{theorem}\label{RTM.th}
Given the two-hop MIMO relay network described by eqs.\,\eqref{ch.eq} with average source and relay power constraints $P_1$ and $P_2$, the optimum (capacity-maximizing) RTM $\Xm$ is given by
\begin{equation}
\Xm=\Umt_B\Lamt_B^{-1/2}\Lamt^{1/2}\Umt_A^\H,
\end{equation}
where the matrices $\Umt_B,\Lamt_B,\Umt_A$ are obtained by the ``thin'' unitary diagonalizations (UD's)~\cite[Th.\ 7.3.2]{horn}:%
\footnote{
A ``thin'' UD $\Um\Lam\Um^\H$ of an $n\times n$ matrix is characterized by an $m\times m$ diagonal matrix $\Lam$ whose diagonal entries are sorted in nonincreasing order, i.e., $(\Lam)_{i,i}\ge(\Lam)_{i+1,i+1}$ for $i=1,\dots,m-1$ and a semi-unitary $n\times m$ matrix $\Um$ with the property that $\Um^\H\Um=\Id_m$.
}
\begin{equation}
\Am=\Umt_A\Lamt_A\Umt_A^\H,\qquad\Bm=\Umt_B\Lamt_B\Umt_B^\H.
\end{equation}
where
\begin{align}\label{ABC.eq}
\begin{array}{ll}
\Am\triangleq\Hm_1\bigg(\dfrac{t}{P_1}\Id_{t}+\Hm_0^\H\Hm_0+\Hm_1^\H\Hm_1\bigg)^{-1}\Hm_1^\H\\
\Bm\triangleq\Hm_2^\H\Hm_2,\qquad
\Cm\triangleq\Id_s+\dfrac{P_1}{t}\Hm_1\Hm_1^\H
\end{array}
\end{align}
We also have
\begin{equation}\label{Lamt.eq}
\Lamt\triangleq\diag(x_1,\dots,x_{\rho},\underbrace{0,\ldots,0}_{\rho_B-\rho})
\end{equation}
where $\rho\triangleq\min(s,\rho_B)$ and $\rho_B\triangleq\rank(\Bm)=\rank(\Hm_2)\le\min(u,r)$.
The diagonal matrix $\Lamt_A$ is possibly extended by zero padding to the size $\rho_B\times\rho_B$.
The matrix $\Lamt$ has $\rho\le\rho_B$ possibly positive eigenvalues, obtained by solving the convex optimization problem
\begin{align}\label{opt1.eq}
\left\{\begin{array}{rl}
\dst\min_{\xv\ge\0}&\dst-\sum_{i=1}^{\rho}\ln\bigg\{1-\frac{\alpha_i}{1+x_i}\bigg\}\\
\mathrm{s.t.}&\dst\sum_{i=1}^{\rho}\beta_ix_i\le P_2,~~x_i\ge0,i=1,\dots,\rho
\end{array}\right.
\end{align}
where, for $i=1,\dots,\rho$,
\begin{align}
\alpha_i&\triangleq(\Lamt_A)_{i,i},&\beta_i&\triangleq\frac{(\Umt_A^\H\Cm\Umt_A)_{i,i}}{(\Lamt_B)_{i,i}}
\end{align}
\end{theorem}
\begin{IEEEproof}
See App. \ref{RTM.app}.
\end{IEEEproof}

\subsection{Parametric Water-Filling solution}\label{param.sec}

We can get a closed-form parametric solution of the optimization problem \eqref{opt1.eq} in Theorem \ref{RTM.th} based on a single $\xi>0$.
To this end, we define%
\footnote{
Hereafter, $\{\cdot\}_+\triangleq\max(0,\cdot)$.
}
\begin{align}\label{x.rho.eq}
\varphi_i(\xi)\triangleq\bigg\{\frac{\alpha_i}{2}-1+\sqrt{\frac{\alpha_i^2}{4}+\frac{\alpha_i}{\beta_i}\xi}\bigg\}_+,~i=1,\dots,\rho.
\end{align}
These functions provide the components of the vector $\xv$, solution of the optimization problem \eqref{opt1.eq} in Theorem \ref{RTM.th}, as $x_i=\varphi_i(\xi)$.
Accordingly, we obtain two parametric equations:
\begin{align}
\label{Pr.eq}    P_2=&\sum_{i=1}^{\rho}\beta_i\varphi_i(\xi)\\
\label{C.rho.eq} C  =&\log_2\det\bigg\{\Id_{t}+\frac{P_1}{t}(\Hm_0^\H\Hm_0+\Hm_1^\H\Hm_1)\bigg\}\non
&+\sum_{i=1}^{\rho}\log_2\bigg\{1-\frac{\alpha_i}{1+\varphi_i(\xi)}\bigg\}
\end{align}
These expressions are obtained by solving the KKT equations corresponding to the optimization problem \eqref{opt1.eq} and are derived in detail in App.~\ref{param.app}.

The uniqueness of the solution of~\eqref{Pr.eq} stems from the fact that the functions $\varphi_i(\xi)$ are monotonically increasing for $\xi>\xi_i\triangleq(1-\alpha_i)\beta_i$.
Since $\varphi_i(\xi)=0$ for $\xi\le\xi_i$, we can find solve~\eqref{Pr.eq} numerically by dividing the real positive line $\{\xi:\xi>0\}$ through the sorted thresholds $\xi_i$ and considering over each interval so determined only the positive functions.
This remains nevertheless a nonlinear equation.
The approach recalls the solution of the water-filling equation arising in the case of independent additive Gaussian channels with an overall average power constraint~\cite{cover}.

\subsection{RTM optimization based on OSTBC capacity}\label{RTM.Th2.sec}

Instead of considering the optimization of the RTM to maximize the capacity, one may consider the maximization of the Orthogonal Space--Time Block Coding (OSTBC) capacity, as defined in~\cite{lars}.
This approach has been followed in~\cite{gaz14}.
Unfortunately, it was assumed in~\cite{gaz14} that the matrix $\Hm_1\Hm_1^\H$ has always full rank $s$, which may not be true, for example, if $t<s$, and limits the generality of the result.
For this reason we provide here a derivation of the optimum RTM achieving the OSTBC capacity in the general case.

The OSTBC capacity with symbol rate $R$ of the MIMO relay channel is given by~\cite{lars}:
\begin{align}\label{C.OSTBC.eq}
C_\mathsf{OSTBC}=R\log_2
&\bigg\{1+\frac{P_1}{tR}\tr\Big[\Hm_0^\H\Hm_0+\Hm_1^\H\Xm^\H\Hm_2^\H\non
&(\Id_r+\Hm_2\Xm\Xm^\H\Hm_2^\H)^{-1}\Hm_2\Xm\Hm_1\Big]\bigg\}
\end{align}
The optimum RTM (maximizing the above capacity for every $R$) is given in the following Theorem.
\begin{theorem}\label{RTM2.th}
Given the two-hop MIMO relay network described by eqs.\,\eqref{ch.eq} with average source and relay power constraints $P_1$ and $P_2$, the optimum (OSTBC capacity-maximizing) RTM $\Xm$ is given by
\begin{equation}
\Xm=\Umt_B\Lamt_B^{-1/2}\Lamt^{1/2}\Umt_A^\H,
\end{equation}
where the matrices $\Umt_B,\Lamt_B,\Umt_A$ are obtained by the ``thin'' unitary UD's
\begin{equation}
\Amc=\Umt_A\Lamt_A\Umt_A^\H,\qquad\Bmc=\Umt_B\Lamt_B\Umt_B^\H.
\end{equation}
where
\begin{align}\label{ABC2.eq}
\Amc&\triangleq\Hm_1\Hm_1^\H,&
\Bmc&\triangleq\Hm_2^\H\Hm_2,&
\Cm\triangleq\Id_s+\frac{P_1}{t}\Hm_1\Hm_1^\H
\end{align}
We define $\Lamt$ as in \eqref{Lamt.eq}, $\rho\triangleq\min(s,\rho_B)$ and $\rho_B\triangleq\rank(\Bm)=\rank(\Hm_2)\le\min(u,r)$.
$\Lamt_A$ is possibly extended by zero padding to the size $\rho_B\times\rho_B$.
The $x_i,i=1,\dots,\rho$ are obtained by solving the optimization problem
\begin{align}\label{opt2.eq}
\left\{\begin{array}{rl}
\dst\min_{\xv\ge\0}&\dst\sum_{i=1}^{\rho}\frac{\alpha_i}{1+x_i}\\
\mathrm{s.t.}&\dst\sum_{i=1}^{\rho}\beta_ix_i\le P_2,~~x_i\ge0,i=1,\dots,\rho
\end{array}\right.
\end{align}
where, for $i=1,\dots,\rho$,
\begin{align}
\alpha_i&\triangleq(\Lamt_A)_{i,i},&\beta_i&\triangleq\frac{(\Umt_A^\H\Cm\Umt_A)_{i,i}}{(\Lamt_B)_{i,i}}
\end{align}
\end{theorem}
\begin{IEEEproof}
See App.\ \ref{RTM2.app}.
\end{IEEEproof}

\subsection{Parametric Water-Filling solution}\label{param2.sec}

Here we provide a closed-form parametric solution to the optimization problem considered in Theorem \ref{RTM2.th}, based on an independent positive variable $\xi$.
Using the definitions of Theorem \ref{RTM2.th}, we define
\begin{align}
\psi_i(\xi)\triangleq\bigg\{\xi\sqrt{\frac{\alpha_i}{\beta_i}}-1\bigg\}_+
\end{align}
Accordingly, we obtain these two parametric equations:
\begin{align}
P_2&=\sum_{i=1}^{\rho}\beta_i\psi_i(\xi)\\
C  &=\log_2\bigg\{1+\frac{P_1}{t}\bigg(\Hm_0^\H\Hm_0+\Hm_1^\H\Hm_1-\sum_{i=1}^{\rho_A}\frac{(\Lam_A)_{i,i}}{1+\psi_i(\xi)}\bigg)\bigg\}
\end{align}
These expressions are derived in detail in App.~\ref{param2.app}.

\section{Numerical results}\label{results.sec}

The numerical results in this section are presented to validate the algorithms derived in cases already handled in the literature and to show their applicability to cases where the literature algorithms are not applicable.

\subsection{Validation of the results}\label{valid.sec}

Here, we compare our algorithms with the results presented by Tang and Hua in their paper, specifically~\cite[Figs.\ 3 and 4]{tang07}.
In that use case, the authors assumed that all the antenna numbers are the same, i.e., $t=r=u=s=M=4$, and considered the ergodic capacity corresponding to a relay system whose channel matrices have all iid Rayleigh distributed fading gains with unit variance without a direct link from source to destination (more precisely, the entries of $\Hm_1,\Hm_2$ are iid $\CN(0,1)$ and $\Hm_0\equiv\0$).
We resort to the following definitions of SNR's:
\begin{equation}
\rho_1\triangleq\frac{P_1}{M\sigma_1^2},\quad
\rho_2\triangleq\frac{P_2}{M\sigma_2^2}.
\end{equation}
According to the previous assumptions, we can simplify the expression of the relay network capacity \eqref{C.eq} as
\begin{align}\label{C.eq2}
C=
&\max_{\Xmc:\tr\{\Xmc(\Id_M+\rho_1\Hm_1\Hm_1^\H)\Xmc^\H\}=M\rho_1}\non
&\hspace*{1cm}\log_2\det\bigg\{\Id_M+\rho_2\Hm_1^\H\Xmc^\H\Hm_2^\H\non
&\hspace*{1cm}\bigg(\Id_M+\frac{\rho_2}{\rho_1}\Hm_2\Xmc\Xmc^\H\Hm_2^\H\bigg)^{-1}\Hm_2\Xmc\Hm_1\bigg\}
\end{align}
Similarly, the OSTBC relay network capacity \eqref{C.OSTBC.eq} becomes
\begin{align}\label{C.OSTBC.eq2}
C_\mathsf{OSTBC}=
&\max_{\Xmc:\tr\{\Xmc(\Id_M+\rho_1\Hm_1\Hm_1^\H)\Xmc^\H\}=M\rho_1}\non
&\log_2\bigg\{1+\tr\bigg[\rho_2\Hm_1^\H\Xmc^\H\Hm_2^\H\non
&\bigg(\Id_M+\frac{\rho_2}{\rho_1}\Hm_2\Xmc\Xmc^\H\Hm_2^\H\bigg)^{-1}\Hm_2\Xmc\Hm_1\bigg]\bigg\}
\end{align}
In both cases, we set $\Xmc\triangleq\sqrt{P_2/P_1}\Xm$ so that the capacity expressions are independent of $P_1,P_2$ and depend only on $\rho_1,\rho_2$.

For this scenario,
Fig.\ \ref{fig3_tang07_new.fig} illustrates the ergodic capacity vs.\ $\rho_1$ at fixed $\rho_2=10$ dB and
Fig.\ \ref{fig4_tang07_new.fig} illustrates the ergodic capacity vs.\ $\rho_2$ at fixed $\rho_1=10$ dB.
Each figure reports the six curves with a label composed of two tags:
the first tag denotes the type of RTM used ({OPT1,OPT2,NAF}) and the second tag denotes the type of capacity plotted (ergodic capacity \eqref{C.eq2} or ergodic OSTBC capacity \eqref{C.OSTBC.eq2}).
The types of RTM's considered are:
$i)$ {OPT1}: RTM maximizing the relay network capacity \eqref{C.eq};
$ii)$ {OPT2}: RTM maximizing the OSTBC capacity \eqref{C.OSTBC.eq};
$iii)$ {NAF}: Naive Amplify and Forward, where the RTM is a scaled identity matrix.
The results agree exactly with those reported in \cite[Figs.\ 3 and 4]{tang07}.

The performances illustrated in Figs.\ \ref{fig3_tang07_new.fig} and \ref{fig4_tang07_new.fig} agree with the basic expectations.
The {OPT1} RTM maximizes the ergodic capacity and is suboptimal for the OSTBC capacity.
The {OPT2} RTM maximizes the ergodic OSTBC capacity and is suboptimal for the capacity.
The {NAF} RTM has the worst performance in all cases.
It is quite noticeable from Fig.\ \ref{fig3_tang07_new.fig} that the OPT2 capacity curve is strongly degraded at large $\rho_1$, as a consequence of the profound mismatch between the capacity and the OSTBC capacity.

As far the asymptotic behavior of the capacity curves in Figs.\ \ref{fig3_tang07_new.fig} and \ref{fig4_tang07_new.fig} is concerned with, we notice a key difference.
In Fig.\ \ref{fig3_tang07_new.fig}, when $\rho_1\to\infty$, all curves converge to different limits, which depend on the fixed value of $\rho_2$.
On the contrary, in Fig.\ \ref{fig4_tang07_new.fig}, when $\rho_2\to\infty$, all the capacity and OSTBC capacity curves converge to the same limits, respectively, which depend on the fixed value of $\rho_1$.
The difference is consistent with the following information theoretical interpretation.
The relay network is the cascade of two channels, channel 1 (source to relay) and channel 2 (relay to destination).
As such, the data-processing inequality \cite{cover}  must be satisfied and the overall capacity is upper bounded by the capacity of each channel.
In the case illustrated in Fig.\ \ref{fig3_tang07_new.fig}, when $\rho_1\to\infty$, channel 1's capacity increases without bound so that the relay network capacity coincides with that of channel 2, and is affected by the RTM.
Hence, the different limits.
As far as OSTBC-capacity maximization is concerned, when $\rho_1\to\infty$, we can see that the RTM rank tends to $1$, so that the capacity and the OSTBC-capacity have the following limiting behavior:
\begin{align}
C&\to\max_{\Xmc}\log_2\det\{\Id_M+\rho_2\Hm_1^\H\Xmc^\H\Hm_2^\H\Hm_2\Xmc\Hm_1\}\\
C_\mathsf{OSTBC}&\to\max_{\Xmc}\log_2\{1+\tr(\rho_2\Hm_1^\H\Xmc^\H\Hm_2^\H\Hm_2\Xmc\Hm_1)\}
\end{align}
Thus, under the limit power constraint $\tr\{\Xmc\Hm_1\Hm_1^\H\Xmc^\H\}=M$, they tend to the same limit.

On the contrary, in the case illustrated in Fig.\ \ref{fig4_tang07_new.fig}, when $\rho_2\to\infty$, channel 2's capacity goes to infinity so that the relay network capacity coincides with that of channel 1, which is independent of the RTM.
Hence, the coincidence of the limits.
Moreover, the upper ergodic capacity limits in both figures coincide.

Finally, we notice that the ergodic OSTBC capacity always entails a major loss (even in the full-rate case) with respect to the ergodic capacity.

\insertfig{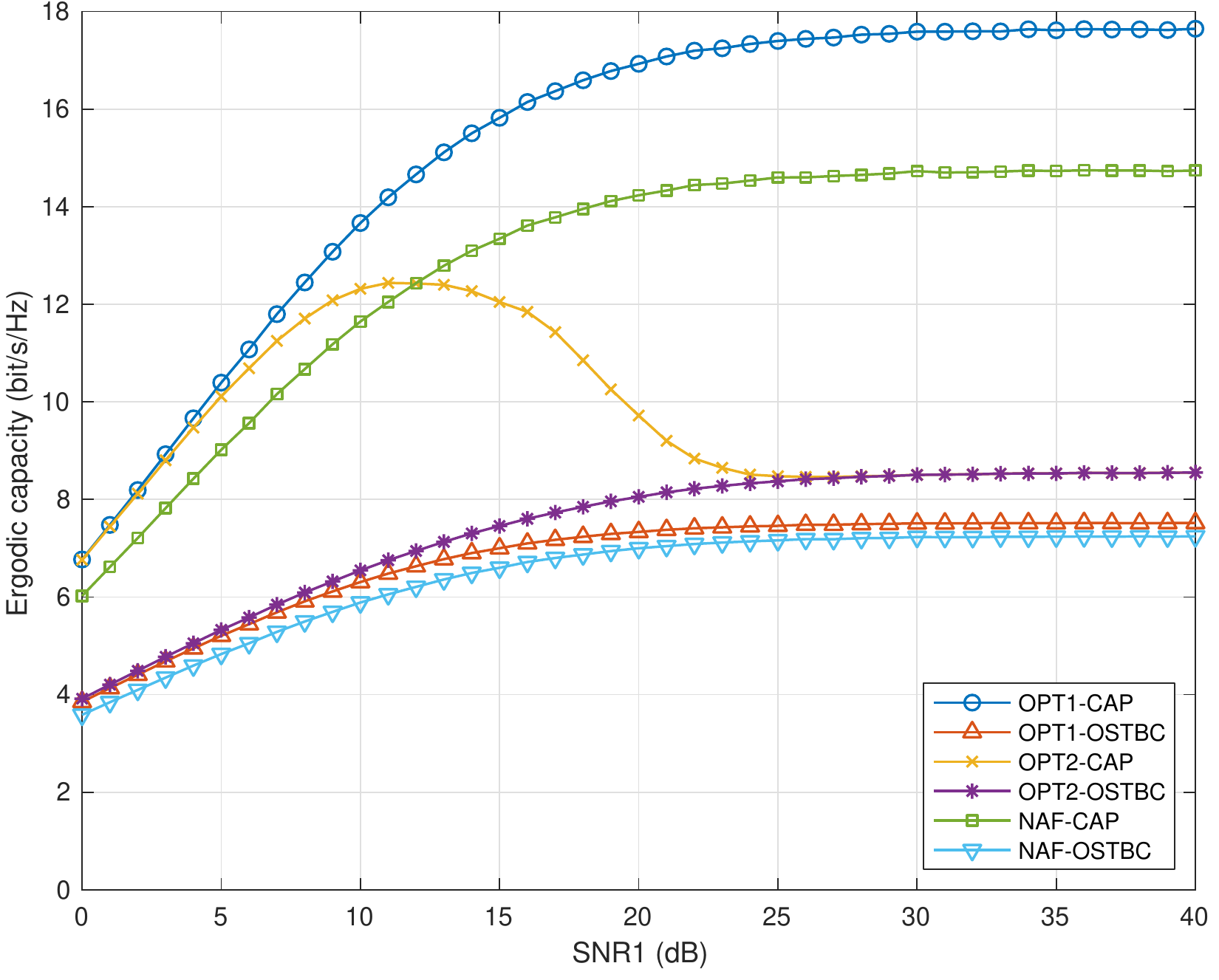}{
Plot of the ergodic capacity vs.\ $\rho_1$ (denoted by SNR1) with $\rho_2=10$ dB, iid Rayleigh fading and three types of RTM.
$i)$ OPT1: optimum RTM for capacity.
$ii)$ OPT2: optimum RTM for full-rate OSTBC capacity.
$iii)$ NAF: Naive Amplify and Forward, the RTM is a scaled identity matrix.
}
\insertfig{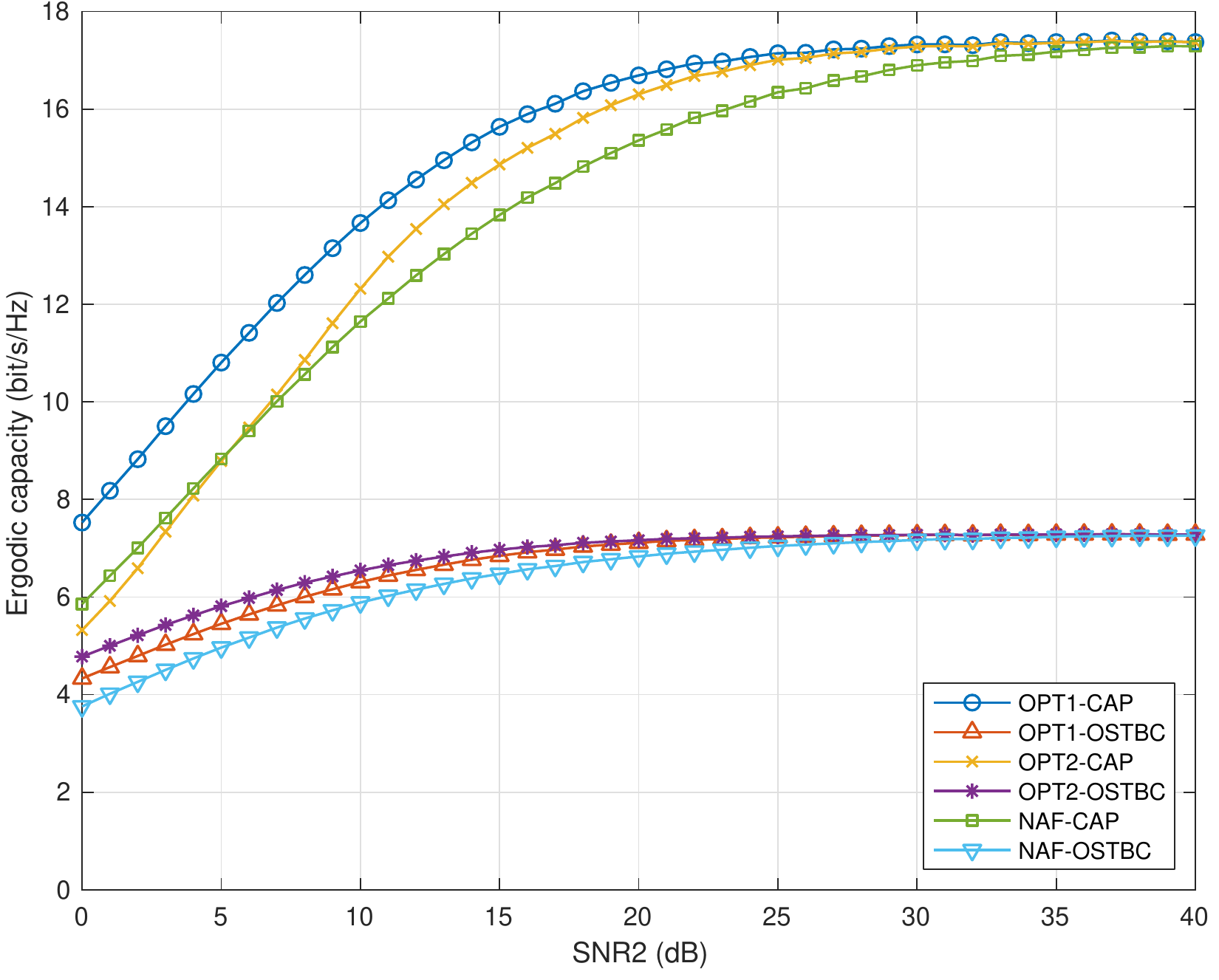}{
Plot of the ergodic capacity vs.\ $\rho_2$ (denoted by SNR2) with $\rho_1=10$ dB, iid Rayleigh fading and three types of RTM.
$i)$ OPT1: optimum RTM for capacity.
$ii)$ OPT2: optimum RTM for full-rate OSTBC capacity.
$iii)$ NAF: Naive Amplify and Forward, the RTM is a scaled identity matrix.
}

\subsection{Full Relay Network --- Equal Number of Antennas}\label{eqnum.sec}

In this case we consider a relay network where also the direct link is present, contrary to the scenario considered in Section \ref{valid.sec}.
We still assume that all antenna arrays have the same number of antennas, $t=r=u=s=M=4$, and the channel matrices are iid Rayleigh as before (i.e., all entries of $\Hm_0,\Hm_1,\Hm_2$ are uncorrelated $\CN(0,1)$ distributed).
Finally, and we define the SNR's as
\begin{equation}
\rho_0\triangleq\frac{P_1}{M\sigma_2^2},\quad
\rho_1\triangleq\frac{P_1}{M\sigma_1^2},\quad
\rho_2\triangleq\frac{P_2}{M\sigma_2^2}.
\end{equation}
Thus, we can simplify the relay network capacity \eqref{C.eq} as:
\begin{align}\label{C.eq3}
C=
&\max_{\Xmc:\tr\{\Xmc(\Id_M+\rho_1\Hm_1\Hm_1^\H)\Xmc^\H\}=M\rho_1}\non
&\hspace*{1cm}\log_2\det\bigg\{\Id_M+\rho_0\Hm_0^\H\Hm_0+\rho_2\Hm_1^\H\Xmc^\H\Hm_2^\H\non
&\hspace*{1cm}\bigg(\Id_M+\frac{\rho_2}{\rho_1}\Hm_2\Xmc\Xmc^\H\Hm_2^\H\bigg)^{-1}\Hm_2\Xmc\Hm_1\bigg\}
\end{align}
Similarly, the OSTBC relay network capacity \eqref{C.OSTBC.eq} becomes
\begin{align}\label{C.OSTBC.eq3}
C_\mathsf{OSTBC}=
&\max_{\Xmc:\tr\{\Xmc(\Id_M+\rho_1\Hm_1\Hm_1^\H)\Xmc^\H\}=M\rho_1}\non
&\log_2\bigg\{1+\tr\bigg[\rho_0\Hm_0^\H\Hm_0+\rho_2\Hm_1^\H\Xmc^\H\Hm_2^\H\non
&\bigg(\Id_M+\frac{\rho_2}{\rho_1}\Hm_2\Xmc\Xmc^\H\Hm_2^\H\bigg)^{-1}\Hm_2\Xmc\Hm_1\bigg]\bigg\}
\end{align}
The two relay network capacity expressions are independent of $P_1,P_2$ for given $\rho_0,\rho_1,\rho_2$.

Fig.\ \ref{fx1_new.fig} illustrates the ergodic capacity behavior vs.\ $\rho_2$ with $\rho_1=10$ dB and two values of $\rho_0=-10,10$~dB.
Again, the optimum for capacity (OPT1), for OSTBC-capacity (OPT2) and naive amplify and forward (NAF) RTM's are considered.
Interestingly, we note that, for low $\rho_0$, \emph{e.g.}, $-10$ dB, the results are very close to those reported in Fig.\ \ref{fig4_tang07_new.fig}.
This condition is close to having no direct link because most power passes through the relay.
Increasing $\rho_0$ impacts drastically on the performance results, as illustrated.
These results allow to assess the trade-offs implied by the presence of the direct link, which is a key contribution of this work.

In a similar way, Fig.\ \ref{fx2_new.fig} plots the ergodic capacity vs.\ $\rho_0$ with fixed  $\rho_1=10$~dB and several values of $\rho_2$: $0,10,20,30$ dB.
In this case, the curves increase monotonically with respect to the link SNR $\rho_0$ and reach a limit as $\rho_2\to\infty$.

By these results we can see when the RTM optimization is worth the effort or rather naive amplify and forward is sufficient for a given scenario.
For example, we can see from Fig.\ \ref{fx2_new.fig} a clear advantage when $\rho_2=10$ dB, which decreases progressively by increasing $\rho_2$, until it becomes very small for $\rho_2=30$ dB.
Then, if the relay-to-destination SNR $\rho_2$ is very large, there is little gain available from RTM optimization, while the gain is substantial in the range of moderate values as $10$ dB.

\insertfig{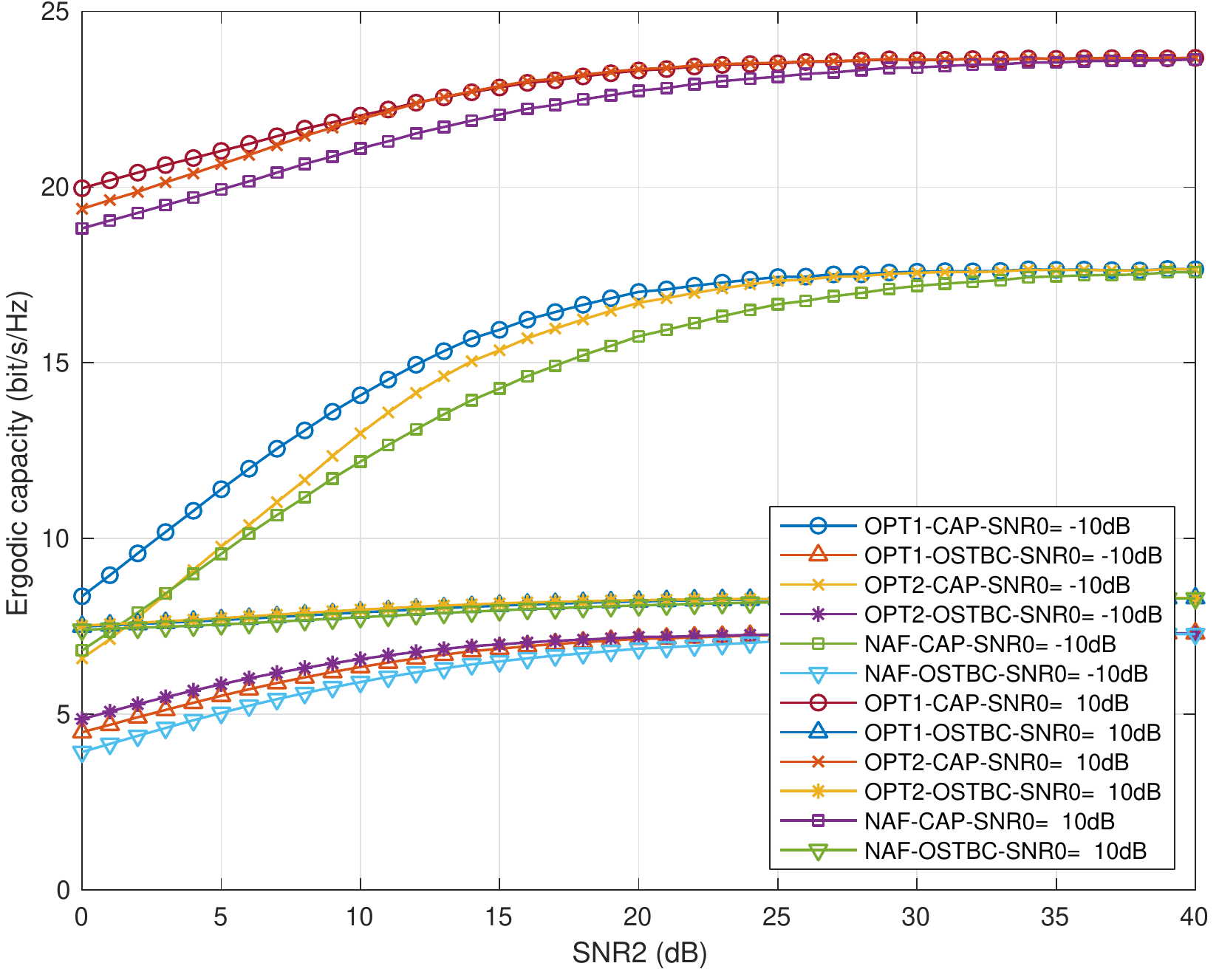}{
Plot of the ergodic capacity vs.\ $\rho_2$ (denoted SNR2) with $\rho_0=-10,10$ dB, $\rho_1=10$ dB, iid Rayleigh fading and three types of RTM.
$i)$ OPT1: optimum RTM for capacity.
$ii)$ OPT2: optimum RTM for full-rate OSTBC capacity.
$iii)$ NAF: Naive Amplify and Forward, the RTM is a scaled identity matrix.
}
\insertfig{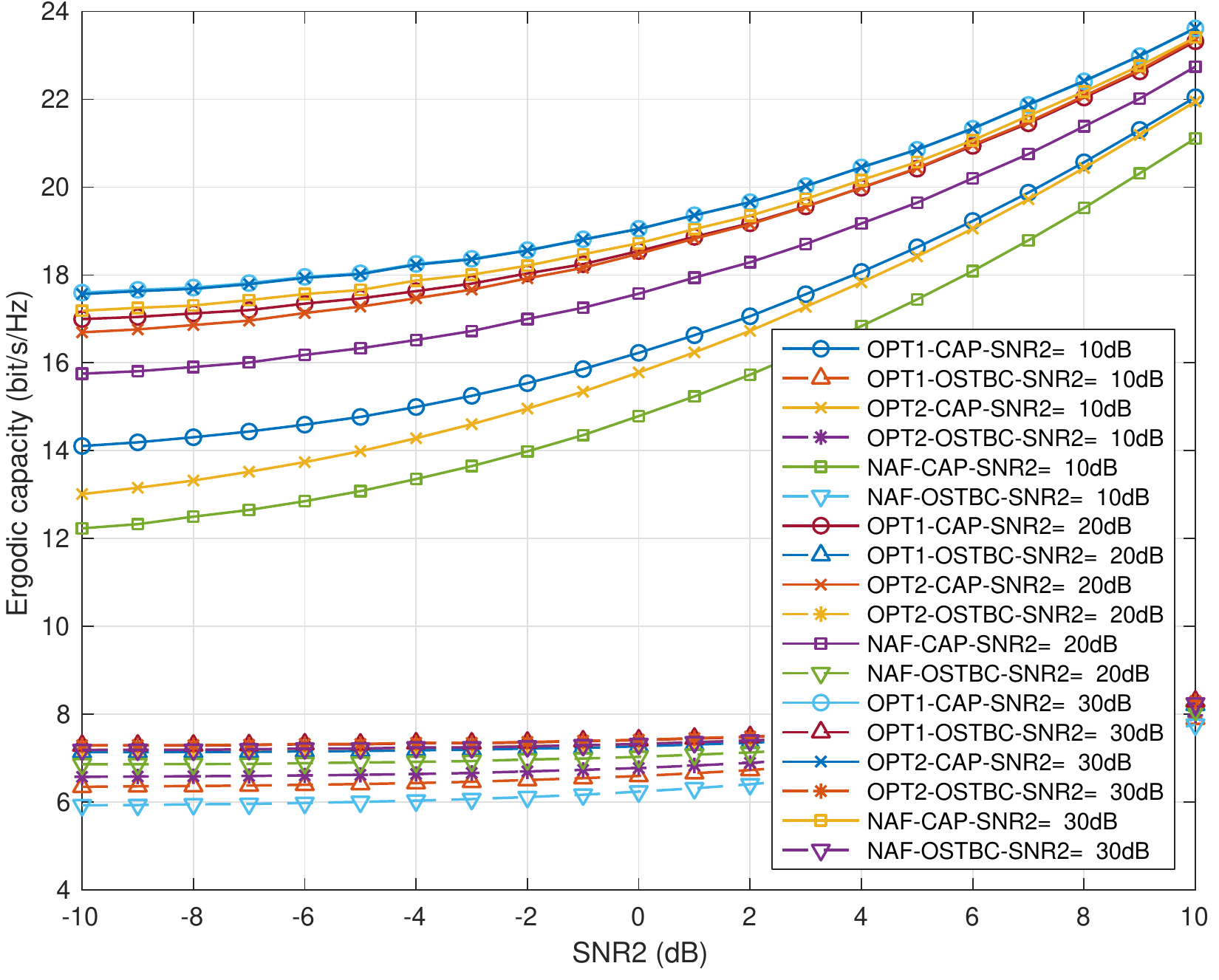}{
Plot of the ergodic capacity vs.\ $\rho_0$ (denoted SNR0) with $\rho_2=0,10,20,30$ dB, $\rho_1=10$ dB, iid Rayleigh fading and three types of RTM.
$i)$ OPT1: optimum RTM for capacity.
$ii)$ OPT2: optimum RTM for full-rate OSTBC capacity.
$iii)$ NAF: Naive Amplify and Forward, the RTM is a scaled identity matrix.
}

\subsection{Full Relay Network --- Different Number of Antennas}\label{varnum.sec}

To conclude this selection of simulation scenarios we consider the case when the number of antennas in the relay network is variable so that the results of the literature are not applicable both for the capacity \cite{tang07} and for the OSTBC-capacity \cite{gaz14}.
In particular, we consider the scenario where the number of transmit and receive antennas of the source and destination are $t=r=2$ and the number of transmit and receive antennas at the relay are $s=u=2$ or $4$ or $8$.
Here, we define
\begin{equation}
\rho_0\triangleq\frac{P_1}{t\sigma_2^2},\quad
\rho_1\triangleq\frac{P_1}{t\sigma_1^2},\quad
\rho_2\triangleq\frac{P_2}{u\sigma_2^2}.
\end{equation}
Figs.\ \ref{fx3R2_new.fig} to \ref{fx3R8_new.fig} show the ergodic capacity of this relay network vs.\ $\rho_2$ with $\rho_1=10$ dB and two values of $\rho_0=-10,10$~dB.
We can see that increasing the number of relay antennas is quite beneficial to the relay network.
In fact, the limit ergodic capacity with $\rho_0=\rho_1=10$ dB and $\rho_2\to\infty$ increases from $9.9$ to $11.4$ and $13.0$ bit/s/Hz as the number of relay antennas increases from $s=u=2$ to $4$ and $8$, respectively, while the number of transmit and receive antennas at the source and destination remain fixed and equal to $2$.
Comparatively, the capacity of the direct link without the relay for $\rho_0=10$ dB is $7.14$ bit/s/Hz~\cite{tel99}.
These results show the effectiveness of a MIMO relay with different numbers of antennas on the capacity.

\insertfig{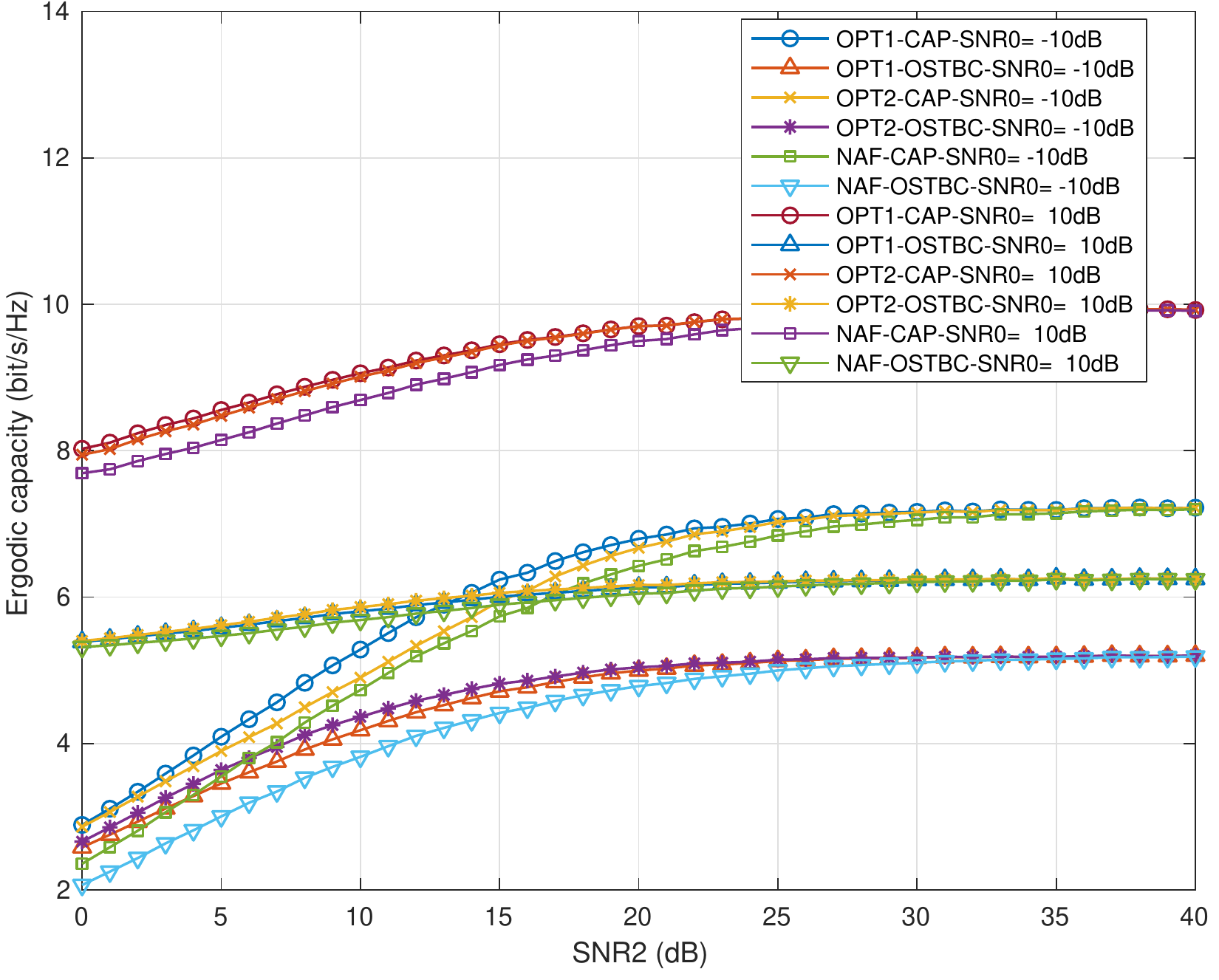}{
Plot of the ergodic capacity vs.\ $\rho_2$ (denoted SNR2) with $\rho_0=-10,10$ dB, $\rho_1=10$ dB, iid Rayleigh fading,
relay network with $t=r=2$ and $s=u=2$, and three types of RTM.
$i)$ OPT1: optimum RTM for capacity.
$ii)$ OPT2: optimum RTM for full-rate OSTBC capacity.
$iii)$ NAF: Naive Amplify and Forward, the RTM is a scaled identity matrix.
}
\insertfig{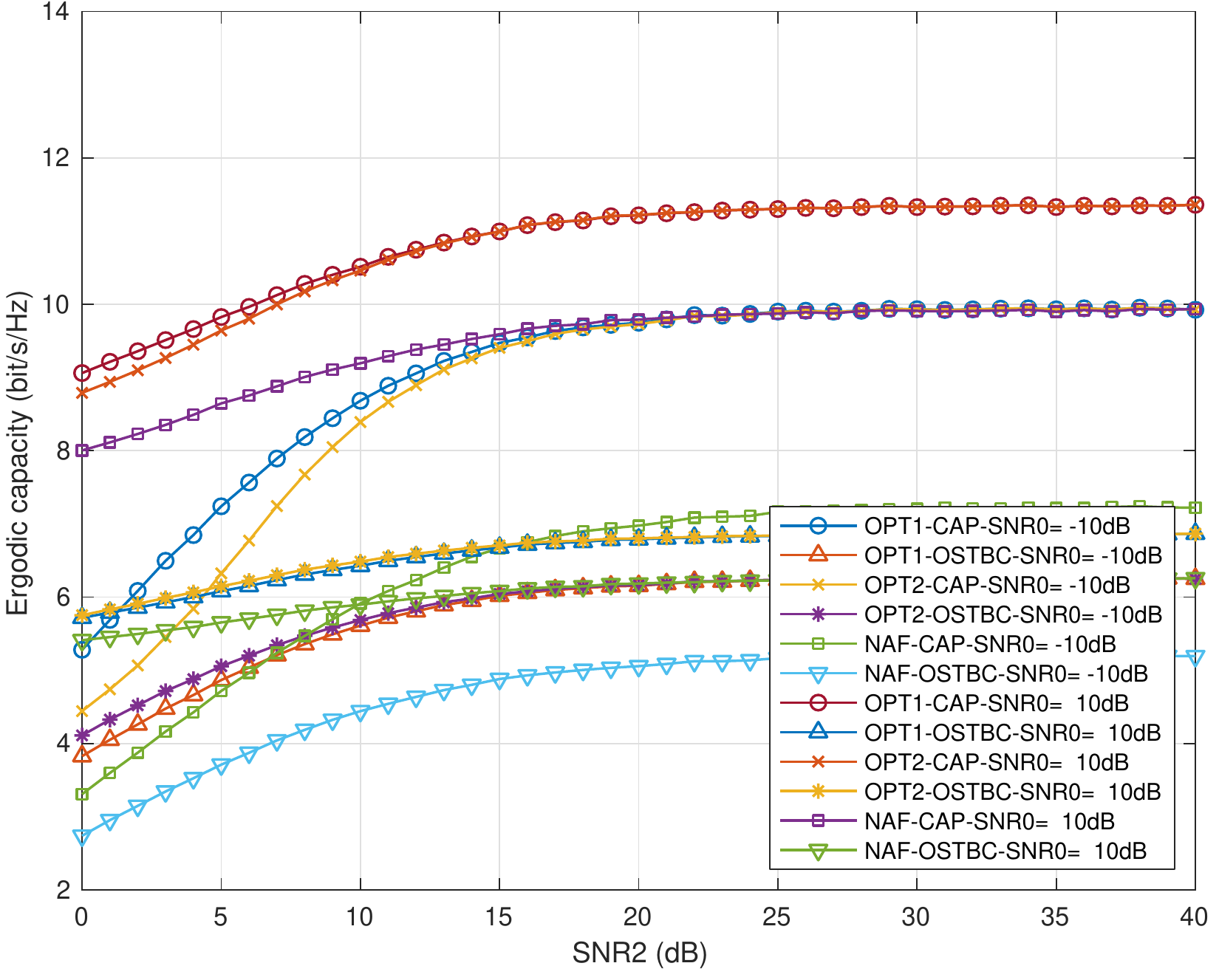}{Same as Fig.\ \ref{fx3R2_new.fig} but $s=u=4$.}
\insertfig{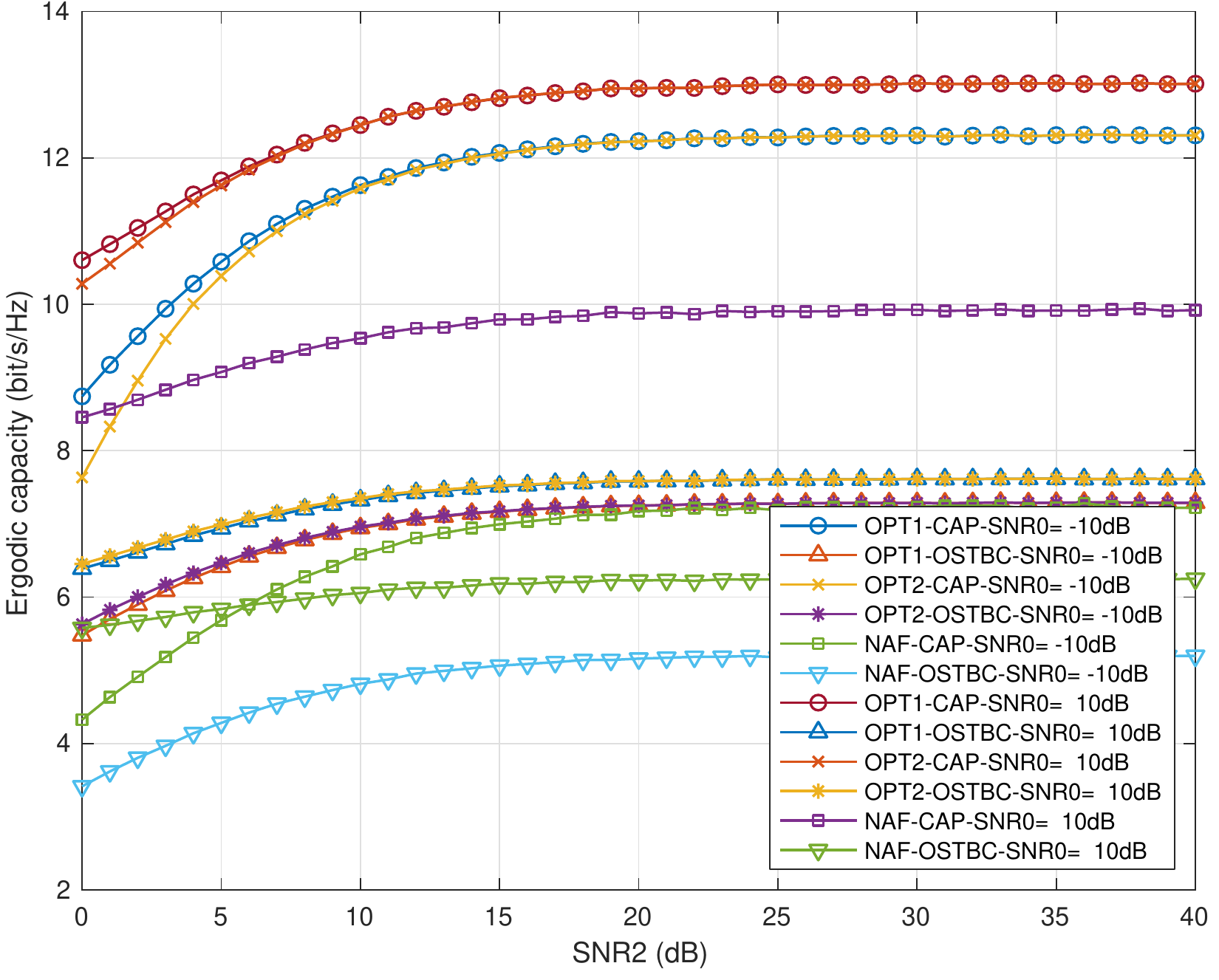}{Same as Fig.\ \ref{fx3R2_new.fig} but $s=u=8$.}

\section{Conclusions}\label{conclusions.sec}

The work focuses on the optimization of the Relay Transformation Matrix (RTM) in a two-hop amplify-and-forward relay network.
The contributions extend nontrivially earlier results from the literature.
The seminal work by Tang and Hua~\cite{tang07} provided the solution of the optimization problem with the capacity as objective function for a pure relay network (without the direct link).
The authors emphasized that the relay network with a direct link case was an open problem at the time and to the author's knowledge it remained so until now.
The work by Shariat and Gazor~\cite{gaz14} established the interest in the full relay network but focused on the OSTBC-capacity only.
Though the OSTBC-optimized RTM provides good results in terms of capacity in many cases, it remains a suboptimal approach and may lead sometimes to considerable performance degradation (see Fig.\ \ref{fig3_tang07_new.fig}).
Moreover, reference~\cite{gaz14} imposed some conditions on the number of antennas of the relay network limiting the generality of the results.
These limitations are overcome in this work which does not assume any conditions on the channel matrices' ranks and on the number of antennas.

The optimum RTM has been derived in Theorems \ref{RTM.th} and \ref{RTM2.th} for the capacity and OSTBC-capacity, respectively, by different simplified convex optimization problems, whose parametric solutions have been derived in Sections \ref{param.sec} and \ref{param2.sec}, respectively.
The KKT equations corresponding to the relevant optimization problems have been solved and used to provide parametric expressions of the average power constraint and the capacity as depending only on a single parameter.
The solution recalls the structure of \emph{water-filling} equations.

Simulation results have been presented to compare the capacity achieved by the optimum RTM and by naive amplify-and-forward.
It is shown that the capacity advantage due to RTM optimization decreases as the SNR increases but it is still sizable for practical SNR values.
To assess the effectiveness of a MIMO relay on an existing $2\times2$ MIMO link we compared different simulation scenarios corresponding to increasing numbers of relay antennas.
For example, we showed in Section~\ref{varnum.sec} that capacity increases from $7.1$~bit/s/Hz (w/o relay) to $9.9$, $11.4$, and $13.0$~bit/s/Hz, by using a relay with $2,4,$ and $8$ antennas, respectively.

\appendices
\section{Proof of Theorem \ref{RTM.th}}\label{RTM.app}

\begin{IEEEproof}
In order to prove the statement of Theorem \ref{RTM.th}, we begin with the following elementary linear algebra identity:
\begin{align}\label{KK.eq}
\Km^\H(\Id+\Km\Km^\H)^{-1}\Km
&=\Km^\H\Km(\Id+\Km^\H\Km)^{-1}\non
&=\Id-(\Id+\Km^\H\Km)^{-1}.
\end{align}
Setting $\Km=\Hm_2\Xm$ in \eqref{KK.eq}, we can rewrite the capacity \eqref{C.eq} as
\begin{align}\label{C.eq1}
C=\log_2
&\det\bigg\{\Id_{t}+
\frac{P_1}{t}(\Hm_0^\H\Hm_0+\Hm_1^\H\Hm_1)\non
&-\frac{P_1}{t}\Hm_1^\H(\Id_s+\Xm^\H\Hm_2^\H\Hm_2\Xm)^{-1}\Hm_1\bigg\}
\end{align}
According to the definition given in Section \ref{RTM.sec}, the optimum RTM is the matrix $\Xm$ that maximizes the capacity reported in eq.~\eqref{C.eq}, under the constraint given by the previous eq.~\eqref{constr.eq}.
Subtracting from eq.~\eqref{C.eq1} the constant term (with respect to $\Xm$)
\begin{equation}
\log_2\det\bigg\{\Id_{t}+\frac{P_1}{t}(\Hm_0^\H\Hm_0+\Hm_1^\H\Hm_1)\bigg\},
\end{equation}
we can see that the optimum RTM is found by solving the following optimization problem:
\begin{align}
\max_{\Xm}\quad  &\det[\Id_t-\Am^\H(\Id_s+\Xm^\H\Bm\Xm)^{-1}\Am]\\
\mathrm{s.t.}\quad &\tr(\Xm\Cm\Xm^\H)\le P_2
\end{align}
where we defined the matrices $\Am,\Bm,\Cm$ as in \eqref{ABC.eq}.
Now, consider the following UD's:
\begin{align}
\label{A.ud.eq}\Am&=\Um_A\Lam_A\Um_A^\H\\
\label{B.ud.eq}\Xm^\H\Bm\Xm&=\Um\Lam\Um^\H
\end{align}
The objective function can be upper bounded as follows:
\begin{align}\label{det.bound}
\det[\Id_{t}-\Am^\H&(\Id_s+\Xm^\H\Bm\Xm)^{-1}\Am]\non
&=\det[\Id_s-\Um_A\Lam_A\Um_A^\H(\Id_s+\Um\Lam\Um^\H)^{-1}]\non
&=\det[\Id_s-\Um^\H\Um_A\Lam_A\Um_A^\H\Um(\Id_s+\Lam)^{-1}]\non
&=\det[\Id_s-\Qm\Lam_A\Qm^\H(\Id_s+\Lam)^{-1}]\non
&=\frac{\det(\Id_s+\Lam-\Qm\Lam_A\Qm^\H)}{\det(\Id_s+\Lam)}\non
&\le\prod_{i=1}^{s}\bigg\{1-\frac{(\Lam_A)_{i,i}}{1+(\Lam)_{i,i}}\bigg\}.
\end{align}
Here, we set $\Qm\triangleq\Um^\H\Um_A$ (i.e., a unitary matrix) and then we applied \cite[eq.(2)]{fied71} after noticing that both $\Lam$ and $\Id_s-\Qm\Lam_A\Qm^\H$ are Hermitian positive semidefinite matrices and the nondecreasingly ordered eigenvalues of $\Id_s-\Qm\Lam_A\Qm^\H$ are $1-(\Lam_A)_{i,i},i=1,\dots,s$.
The upper bound is attained by setting $\Qm=\Id_s$.
Hence, $\Um=\Um_A$.

To find an expression of the RTM $\Xm$, we notice that both sides of \eqref{B.ud.eq} have the same rank:
\begin{equation}
\rho\triangleq\rank(\Xm^\H\Bm\Xm)=\rank(\Um_A\Lam\Um_A^\H)\le\min(s,u,r).
\end{equation}
If $\rho_B\triangleq\rank(\Bm)$, then $\rho_B\ge\rho$, and we have the following ``thin'' UD's:
\begin{equation}
\Bm=\underbrace{\Umt_B\Lamt_B\Umt_B^\H}_{u\times\rho_B\times\rho_B\times u}\qquad
\Um_A\Lam\Um_A^\Hm=\underbrace{\Umt_A\Lamt\Umt_A^\H}_{s\times\rho_B\times\rho_B\times s}
\end{equation}
where $\Umt_B^\H\Umt_B=\Umt_A^\H\Umt_A=\Id_{\rho_B}$,
$\Lamt_B$ is the diagonal submatrix of $\Lam_B$ with the positive elements, and
$\Lamt$ is the \emph{unknown} diagonal submatrix of $\Lam$ with nonnegative elements while the other elements of $\Lam$ (if any) are all equal to $0$.
Thus, eq.\ \eqref{B.ud.eq} is satisfied by setting
\begin{equation}\label{Xm.eq}
\Xm=\Umt_B\Lamt_B^{-1/2}\Lamt^{1/2}\Umt_A^\H.
\end{equation}

\begin{remark}
Notice that the maximum in \eqref{det.bound} is attained regardless of any constraint by the matrix $\Xm$ with the structure given in \eqref{Xm.eq}.
For every pair of Hermitian positive semidefinite matrices $\Am$ and $\Bm$, the matrix defined in \eqref{Xm.eq} maximizes
\begin{equation}
\det[\Id_{t}-\Am^\H(\Id_s+\Xm^\H\Bm\Xm)^{-1}\Am],
\end{equation}
and thus the capacity \eqref{C.eq1}.
The structure \eqref{Xm.eq} contains $\rho$ free parameters as the diagonal elements of $\Lamt$.
The relay power constraint is introduced in the following optimization problem.
\end{remark}

Now, we have to choose $\Lamt$ in order to 
$i)$ maximize the upper bound in \eqref{det.bound}, namely, $$\prod_{i=1}^{s}\bigg\{1-\frac{(\Lam_A)_{i,i}}{1+(\Lam)_{i,i}}\bigg\},$$
and $ii)$ satisfy the relay power constraint \eqref{constr.eq}, which can be written as
\begin{equation}
\sum_{i=1}^{\rho}\frac{(\Umt^\H\Cm\Umt)_{i,i}}{(\Lamt_B)_{i,i}}(\Lam)_{i,i}=P_2.
\end{equation}
Notice that the inequality in \eqref{constr.eq} is turned into an equality since a possibly optimum solution $\Lam_0$ such that
\begin{equation}
\sum_{i=1}^{\rho}\frac{(\Umt^\H\Cm\Umt)_{i,i}}{(\Lamt_B)_{i,i}}(\Lam_0)_{i,i}=\rho P_2<P_2,
\end{equation}
for some $0<\rho<1$,
cannot be optimum since $\rho^{-1}\Lam_0$ would increase all the factors in the upper bound in \eqref{det.bound} since
\begin{equation}
1-\frac{(\Lam_A)_{i,i}}{1+\rho^{-1}(\Lam_0)_{i,i}}>1-\frac{(\Lam_A)_{i,i}}{1+(\Lam_0)_{i,i}}.
\end{equation}

The detailed solution of this optimization problem is reported in the following App.\ \ref{param.app} and completes the proof of Theorem \ref{RTM.th}.
\end{IEEEproof}

\section{Parametric solution of optimization problem \eqref{opt1.eq}}\label{param.app}

From the statement of the optimization problem reported in eq.\ \eqref{opt1.eq} of Theorem \ref{RTM.th}, we derive the Lagrangian function of the problem as follows:
\begin{align}
\L(\xv,&\lambda_0,\lambda_1,\dots,\lambda_\rho)=-\sum_{i=1}^{\rho}
\ln\bigg\{1-\frac{\alpha_i}{1+x_i}\bigg\}\non
&+\lambda_0(\betav^\T\xv-P_2)-\sum_{i=1}^{\rho}\lambda_ix_i.
\end{align}
where $0<\alpha_i<1,\beta_i>0$ and $\lambda_i,i=0,\dots,\rho,$ are the Lagrange multipliers~\cite{boyd}.
Here, we did not consider the constraints $x_1\ge x_2\ge\dots\ge x_\rho$ since, by Lemma \ref{lemma.lem}, these constraint are automatically satisfied by any nonnegative solution ($x_i\ge0,i=1,\dots,\rho$).
In fact, otherwise, a permutation of the variables would lead to a further decrease of the objective function.
Thus, we can save the extra effort that would be required.

The KKT equations are obtained according to \cite[Sec.\,5.5.3]{boyd}.
First, we take the partial derivatives of the Lagrangian function with respect to the variables $x_i$, for $i=1,\dots,\rho$:
\begin{equation}
\frac{\partial\L}{\partial x_i}=\frac{1}{1+x_i}-\frac{1}{1-\alpha_i+x_i}+\lambda_0\beta_i-\lambda_i
\end{equation}
Then, we have the following KKT equations:
\begin{equation}\label{kkt.appb.eq}
\left\{\begin{array}{lll}
\betav^\T\xv-P_2&\le0\\
\lambda_0(\betav^\T\xv-P_2)&=0\\
\lambda_0&\ge0\\
-x_i&\le0&i=1,\dots,\rho\\
\lambda_ix_i&=0&i=1,\dots,\rho\\
\lambda_i&\ge0&i=1,\dots,\rho\\
\dfrac{\partial\L}{\partial x_i}&=0&i=1,\dots,\rho\\
\end{array}\right.
\end{equation}
We can see that the objective function
\begin{equation}
f(\xv)\triangleq-\sum_{i=1}^{\rho}\ln\bigg\{1-\frac{\alpha_i}{1+x_i}\bigg\}
\end{equation}
is convex for $\xv\ge\0$ because
\begin{equation}
\frac{\partial^2f}{\partial x_i^2}=\frac{\alpha_i(2-\alpha_i+2x_i)}{(1+x_i)^2(1-\alpha_i+x_i)^2}\ge0.
\end{equation}
The mixed derivatives $\partial^2f/(\partial x_i\partial x_j)=0$ for all $i\ne j$.
Therefore, we have a convex optimization problem.
We can see that Slater's condition is satisfied, so that the KKT equations are sufficient for optimality.

The constraint $\betav^\T\xv-P_2\le0$ is achieved with equality since $f(\xv)$ is decreasing with every $x_i$.
Therefore, we have $\lambda_0\ge0$.

Finally, we obtain from the gradient equations:
\begin{equation}
\frac{1}{1-\alpha_i+x_i}-\frac{1}{1+x_i}=\lambda_0\beta_i-\lambda_i,\quad i=1,\dots,\rho.
\end{equation}
For a given $\lambda_0\ge0$, recalling that $\lambda_i\ge0,x_i\ge0,\lambda_ix_i=0$, there are two possible cases
\begin{itemize}
\item $\lambda_i=0$, which implies that the equation is equivalent to
\begin{equation}\label{poly2.eq}
x_i^2+(2-\alpha_i)x_i+1-\alpha_i-\frac{\alpha_i}{\lambda_0\beta_i}=0
\end{equation}
Since $0<\alpha_i<1$, a solution $x_i>0$ exists only if
\begin{equation}
1-\alpha_i-\frac{\alpha_i}{\lambda_0\beta_i}<0\Longrightarrow\lambda_0<\frac{\alpha_i}{(1-\alpha_i)\beta_i}
\end{equation}
and is given by
\begin{equation}
x_i=\frac{\alpha_i}{2}-1+\sqrt{\frac{\alpha_i^2}{4}+\frac{\alpha_i}{\lambda_0\beta_i}}.
\end{equation}
\item $\lambda_i>0$, which implies that one root of \eqref{poly2.eq} must be equal to $0$ to satisfy the KKT condition $\lambda_ix_i=0$.
In turn, this implies that
\begin{equation}
1-\alpha_i-\frac{\alpha_i}{\lambda_0\beta_i-\lambda_i}=0
\end{equation}
and hence
\begin{equation}
\lambda_0=\frac{\alpha_i}{(1-\alpha_i)\beta_i}+\frac{\lambda_i}{\beta_i}>\frac{\alpha_i}{(1-\alpha_i)\beta_i},
\end{equation}
so that
\begin{equation}
\frac{\alpha_i}{2}-1+\sqrt{\frac{\alpha_i^2}{4}+\frac{1}{\lambda_0\beta_i}}<
\frac{\alpha_i}{2}-1+\sqrt{\frac{\alpha_i^2}{4}+1-\alpha_i}=0
\end{equation}

\end{itemize}
Summarizing, we can write the solution in all cases as
\begin{equation}
x_i=\bigg\{\frac{\alpha_i}{2}-1+\sqrt{\frac{\alpha_i^2}{4}+\frac{\alpha_i}{\lambda_0\beta_i}}\bigg\}_+
\end{equation}
Thus, the unknown $\lambda_0\ge0$ can be found by solving the nonlinear equation
\begin{equation}
P_2=\sum_{i=1}^{\rho}\beta_i\bigg\{\frac{\alpha_i}{2}-1+\sqrt{\frac{\alpha_i^2}{4}+\frac{\alpha_i}{\lambda_0\beta_i}}\bigg\}_+
\end{equation}
A unique solution always exists because the rhs is a monotonically decreasing function of $\lambda_0$, which is identically equal to $0$ when $\lambda_0\ge\max_{1\le i\le\rho}\frac{\alpha_i}{(1-\alpha_i)\beta_i}$.
Setting $\xi\triangleq1/\lambda_0$ yields the parametric solution reported in eqs.~\eqref{x.rho.eq} to~\eqref{C.rho.eq}.

\section{Proof of Theorem \ref{RTM2.th}}\label{RTM2.app}

\begin{IEEEproof}
We proceed, as in the proof of Theorem \ref{RTM.th} of Appendix \ref{RTM.app}, to apply the identity \eqref{KK.eq} to the trace argument of \eqref{C.OSTBC.eq}.
We obtain:
\begin{align}
&\hspace*{-1cm}
\Hm_1^\H\Xm^\H\Hm_2^\H(\Id_r+\Hm_2\Xm\Xm^\H\Hm_2^\H)^{-1}\Hm_2\Xm\Hm_1\\
&=\Hm_1^\H\Hm_1-\Hm_1^\H(\Id_s+\Xm^\H\Hm_2^\H\Hm_2\Xm)^{-1}\Hm_1
\end{align}
After defining the matrices $\Amc,\Bmc,\Cm$ as in \eqref{ABC2.eq}, we get the following expression for the optimization problem to maximize the OSTBC capacity \eqref{C.OSTBC.eq}:
\begin{equation}\label{opt2app.eq}
\left\{\begin{array}{rl}\displaystyle
\min_{\Xm}~&\tr\{\Amc(\Id_s+\Xm^\H\Bmc\Xm)^{-1}\}\\
\mathrm{s.t.}~&\tr\{\Xm\Cm\Xm^\H\}\le P_2
\end{array}\right.
\end{equation}
Calculating the UD's
\begin{align}
\Amc&=\Umc_A\Lamc_A\Umc_A^\H&\Xm^\H\Bmc\Xm&=\Umc\Lamc\Umc^\H
\end{align}
and defining the matrix $\Qmc\triangleq\Umc^\H\Umc_A$, we can rewrite optimization problem \eqref{opt2app.eq} as
\begin{align}
\left\{\begin{array}{rl}\displaystyle
\min_{\Xm}~&\tr\{(\Id+\Lamc)^{-1}\Qmc\Lamc_A\Qmc^\H\}\\
\mathrm{s.t.}~&\tr\{\Xm\Cm\Xm^\H\}\le P_2
\end{array}\right.
\end{align}
The objective function can be written as
\begin{align}
\tr\{(\Id+\Lamc)^{-1}\Qmc\Lamc_A\Qmc^\H\}
&=\sum_{i=1}^{s}\sum_{j=1}^{s}\frac{|(\Qmc)_{i,j}|^2(\Lamc_A)_{j,j}}{1+(\Lamc)_{i,i}}\\
&=\lamvt^\T\Qmt\lamv_A
\end{align}
where we defined the column vectors $\lamvt,\lamv_A$ and the matrix $\Qmt$ by
\begin{align}
(\lamvt)_i  &\triangleq\frac{1}{1+(\Lamc)_{i,i}} &i=1,\dots,s\\
(\lamv_A)_j &\triangleq(\Lamc_A)_{j,j} &j=1,\dots,s\\
(\Qmt)_{i,j}&\triangleq|(\Qmc)_{i,j}|^2 &i,j=1,\dots,s
\end{align}
Since $\Qm$ is a unitary matrix, $\Qmt$ is a \emph{doubly stochastic} matrix and, by Birkhoff's theorem~\cite[Th.8.7.2]{horn}, it can be written as the weighted sum of a certain number of permutation matrices:
\begin{equation}
\Qmt=\sum_{\ell=1}^{N}w_\ell\Pim_\ell
\end{equation}
with $N\le s^2-s+1$, $w_\ell\ge0,\ell=1,\dots,N$, and $\sum_{\ell=1}^{N}w_\ell=1$.
Thus,
\begin{align}
\min_{\Qmc:\Qmc\Qmc^\H=\Id}\tr\{(\Id+\Lamc)^{-1}&\Qmc\Lamc_A\Qmc^\H\}
 =\sum_{\ell=1}^{N}w_\ell\lamvt^\T\Pim_\ell\lamv_A\non
&=\min_{1\le\ell\le N}\lamvt^\T\Pim_\ell\lamv_A\non
&=\min_{1\le\ell\le N}\sum_{i=1}^{s}\frac{(\Lamc_A)_{\pi_\ell(i),\pi_\ell(i)}}{1+(\Lamc)_{i,i}}
\end{align}
where $\pi_\ell$ is the permutation associated to the permutation matrix $\Pim_\ell$ defined by
\begin{equation}
(\Pim_\ell)_{i,j}=\delta_{\pi_\ell(i),j}
\end{equation}
where $\delta_{a,b}=1$ if $a=b$ and $0$ otherwise (Kronecker delta function).
The optimum permutation can be found by applying the lower bound of Lemma \ref{lemma.lem} from Appendix \ref{lemma.app}:
\begin{equation}
\min_{\Qmc:\Qmc\Qmc^\H=\Id}\tr\{(\Id+\Lamc)^{-1}\Qmc\Lamc_A\Qmc^\H\}=\sum_{i=1}^{\rho_A}\frac{(\Lamc_A)_{i,i}}{1+(\Lamc)_{i,i}}
\end{equation}
where $\rho_A\triangleq\rank(\Amc)=\rank(\Hm_1)\le\min(t,s)$.
We notice that the optimum solution found above corresponds to setting $\Qmc=\Id_s$, which implies $\Umc=\Umc_A$.
We also have to take into account the additional constraint stemming from the inequality
\begin{equation}
\rho\triangleq\rank(\Lamc)\le\min(\rho_B,s)
\end{equation}
where $\rho_B\triangleq\rank(\Bmc)=\rank(\Hm_2)\le\min(r,u)$.
Using the ``thin'' UD
\begin{equation}
\Bmc=\underbrace{\Umt_B\Lamt_B\Umt_B^\H}_{u\times\rho_B\times\rho_B\times u},
\end{equation}
eq.\ \eqref{B.ud.eq} is satisfied by setting
\begin{equation}
\Xm=\Umt_B\Lamt_B^{-1/2}\Lamt^{1/2}\Umt_A^\H,
\end{equation}
where $\Umt_A$ is obtained by taking the first $\rho$ columns of $\Um_A$,
and the relay power constraint \eqref{constr.eq} becomes
\begin{equation}
\sum_{i=1}^{\rho}\frac{(\Umt_A^\H\Cm\Umt_A)_{i,i}}{(\Lamt_B)_{i,i}}(\Lamc)_{i,i}\le P_2,
\end{equation}
which completes the proof of Theorem \ref{RTM2.th}.
\end{IEEEproof}

\section{Parametric solution of optimization problem \eqref{opt2.eq}}\label{param2.app}

We proceed as in App.\ \ref{param.app} with the Lagrangian
\begin{align}
\L(\xv,&\lambda_0,\lambda_1,\dots,\lambda_\rho)=\sum_{i=1}^{\rho}\frac{\alpha_i}{1+x_i}+\lambda_0(\betav^\T\xv-P_2)\non
&-\sum_{i=1}^{\rho}\lambda_ix_i.
\end{align}
The Lagrangian derivatives are
\begin{equation}
\frac{\partial\L}{\partial x_i}=-\frac{\alpha_i}{(1+x_i)^2}+\lambda_0\beta_i-\lambda_i
\end{equation}
The KKT equations remain the same as \eqref{kkt.appb.eq} from App.\ \ref{param.app}.
The objective function is
\begin{equation}
f(\xv)\triangleq\sum_{i=1}^{\rho}\frac{\alpha_i}{1+x_i},
\end{equation}
which is plainly convex for $\xv\ge\0$ so that we have a convex optimization problem.
Slater's condition is satisfied so that the KKT equations are sufficient for optimality.
Again, $\betav^\T\xv-P_2\le0$ is achieved with equality since $f(\xv)$ is decreasing with every $x_i$, so that $\lambda_0\ge0$.
The gradient equations are:
\begin{equation}
\frac{\alpha_i}{(1+x_i)^2}=\lambda_0\beta_i-\lambda_i,\quad i=1,\dots,\rho.
\end{equation}
For a given $\lambda_0\ge0$, recalling that $\lambda_i\ge0,x_i\ge0,\lambda_ix_i=0$, we can get the solution:
\begin{equation}
x_i=\bigg\{\xi\sqrt{\frac{\alpha_i}{\beta_i}}-1\bigg\}_+
\end{equation}
where $\xi\triangleq\lambda_0^{-1/2}$ and $\{\cdot\}_+\triangleq\max(0,\cdot)$.
Thus, the unknown $\xi>0$ can be found by solving the nonlinear equation%
\footnote{$1_{\A}=1$ when $\A$ is true and $0$ otherwise.}
$$P_2=\sum_{i=1}^{\rho}(\xi\sqrt{\alpha_i\beta_i}-\beta_i)\cdot1_{\xi>\sqrt{\beta_i/\alpha_i}}$$
A unique solution always exists because the rhs is a monotonically increasing function of $\xi$ for
$$\xi\ge\min_{1\le i\le\rho}\sqrt{\frac{\beta_i}{\alpha_i}}.$$

\section{Sequence product sum lemma}\label{lemma.app}

\begin{lemma}\label{lemma.lem}
Given any two real nonnegative nonincreasing sequences $\alpha_i,\beta_i,i=1,\dots,n$ such that $\alpha_i\ge\alpha_{i+1}$ and $\beta_i\ge\beta_{i+1}$, for $i=1,\dots,n-1$, we have, for every permutation $\pi$, the following inequality:
\begin{equation}
\sum_{i=1}^{n}\alpha_i\beta_{n+1-i}\le\sum_{i=1}^{n}\alpha_i\beta_{\pi(i)}\le\sum_{i=1}^{n}\alpha_i\beta_i.
\end{equation}
\end{lemma}
\begin{IEEEproof}
Since every permutation $\pi\in S_n$ can be expressed as a product of disjoint \emph{cycles}~\cite[Sec.~III.70]{princeton}, we have to prove the inequalities only when $\pi$ is a cycle and then apply it to any $\pi\in S_n$ after proper relabeling of the indexes.
Let us assume, w.l.o.g., that $\pi=(1,\dots,n)$, i.e., the permutation $1\mapsto2\mapsto3\mapsto\dots\mapsto n\mapsto1$.
For the upper bound, we have to show that
\begin{align*}
\alpha_1(\beta_1-\beta_2)+\alpha_2(\beta_2-\beta_3)+\dots+\alpha_n(\beta_n-\beta_1)\ge0.
\end{align*}
The above inequality stems from the following:
\begin{align*}
&\alpha_1(\beta_1-\beta_2)+\alpha_2(\beta_2-\beta_3)+\dots+\alpha_n(\beta_n-\beta_1)\non
&=(\alpha_1-\alpha_n)(\beta_1-\beta_2)+\dots+(\alpha_{n-1}-\alpha_n)(\beta_{n-1}-\beta_n)\non
&\ge0,
\end{align*}
since $\alpha_i-\alpha_n\ge0$ and $\beta_i-\beta_{i+1}\ge0$ for every $i=1,\dots,n-1$.

Similary, for the lower bound, we have to show that
\begin{align*}
&\alpha_1(\beta_n-\beta_{n-1})+\alpha_2(\beta_{n-1}-\beta_{n-2})\\
&\hspace*{4cm}+\dots+\alpha_n(\beta_1-\beta_n)\le0.
\end{align*}
The above inequality stems from the following:
\begin{align*}
&\alpha_1(\beta_n-\beta_{n-1})+\alpha_2(\beta_{n-1}-\beta_{n-2})+\dots+\alpha_n(\beta_1-\beta_n)\non
&=(\alpha_1-\alpha_n)(\beta_n-\beta_{n-1})+\dots+(\alpha_{n-1}-\alpha_n)(\beta_2-\beta_1)\non
&\le0,
\end{align*}
since $\alpha_i-\alpha_n\ge0$ and $\beta_i-\beta_{i-1}\le0$ for every $i=1,\dots,n-1$.
\end{IEEEproof}



\begin{thebibliography}{XX}

\bibitem{vdm71}
E.C.\ Van Der Meulen, ``Three terminal communication channels,'' \emph{Adv. Appl. Prob.}, vol. 3, pp. 120--154, 1971.

\bibitem{cover79}
T.M.\ Cover and A.A.\ El Gamal, ``Capacity theorems for the relay channel,'' \emph{IEEE Trans. Inf. Theory}, vol. 25, no. 5, pp. 572--584, Sept. 1979.

\bibitem{send03}
A.\ Sendonaris, E.\ Erkip, and B.\ Aazhang, ``User cooperation diversity -- part I and part II,'' \emph{IEEE Trans. Commun.}, vol. 51, no. 11, pp. 1927--1948, Nov. 2003.

\bibitem{nab04}
R.U.\ Nabar, H.\ Bolcskei, and F.W. Kneubuhler, ``Fading relay channels: performance limits and space-time signal design,'' \emph{IEEE J. Sel. Areas Commun.}, vol. 22, no. 6, pp. 1099--1109, Aug. 2004.

\bibitem{lane04}
J.N.\ Laneman, D.N.C.\ Tse, and G.W.\ Wornell, ``Cooperative diversity in wireless networks: efficient protocols and outage behavior,'' \emph{IEEE Trans. Inf. Theory}, vol. 50, no. 12, pp. 3062--3080, Dec. 2004.

\bibitem{host05}
A. Host-Madsen and J. Zhang, “Capacity bounds and power allocation for wireless relay channel,” \emph{IEEE Trans. Inf. Theory}, vol. 51, no. 6, pp. 2020–2040, June 2005.

\bibitem{wang05}
B.\ Wang, J.\ Zhang, and A.\ Host-Madsen, ``On the capacity of MIMO relay channels,'' \emph{IEEE Trans. Inf. Theory}, vol. 51, no. 1, pp. 29--43, Jan. 2005.

\bibitem{tang07}
X. Tang and Y. Hua, “Optimal design of non-regenerative MIMO wireless relays,” IEEE Trans. Wireless Commun., vol. 6, no. 4, pp. 1398–1407, 2007.

\bibitem{gaz14}
M.H.\ Shariat and S.\ Gazor, ``Optimal non-regenerative linear MIMO relay for orthogonal space time codes,'' in \emph{IEEE Signal Processing Letters}, vol.~21, no.~2, pp.~163--167, Feb.~2014.

\bibitem{lee14}
K.C.\ Lee, C.P.\ Li,T.Y.\ Wang, and H.J.\ Li, ``Performance analysis of dual-hop amplify-and-forward systems with multiple antennas and co-channel interference,'' \emph{IEEE Trans. Wireless Commun.}, vol. 13, no. 6, pp. 3070--3087, June 2014.

\bibitem{liang17}
X.\ Liang, Z.\ Ding, and C.\ Xiao, ``On linear precoding of nonregenerative MIMO relay networks for finite-alphabet source,'' \emph{IEEE Trans. Veh. Technol.}, vol. 66, no. 11, pp. 9005--9017, Nov. 2017.

\bibitem{palla1}
Y. Zhang, J. Li, L. Pang, and Z. Ding, “On precoder design for amplify-and-forward MIMO relay systems,” in \emph{Proc. IEEE VTC-Fall}, 2011.
\bibitem{palla2}
W. Park, S. Jeong, H.-Y. Song, and C. Lee, “The global optimality of the MIMO cooperative system with source and relay precoders for capacity maximization,” \emph{IEEE Trans. Commun.}, vol. 60, no. 10, pp. 2886--2892, Oct. 2012.
\bibitem{palla3}
H. Wan and W. Chen, “Joint source and relay design for multiuser MIMO nonregenerative relay networks with direct links,” \emph{IEEE Trans. Veh. Techn.}, vol. 61, no. 6, pp. 2871--2876, July 2012.
\bibitem{palla4}
R. Mo and Y. H. Chew, “Precoder design for non-regenerative MIMO relay systems,” \emph{IEEE Trans. Wireless Commun.}, vol. 8, no. 10, pp. 5041-- 5049, Oct. 2009.
\bibitem{palla5}
R. Zhang, S.-H. Leung, Z. Luo, and H. Wang, “Precoding design for correlated MIMO-AF relay networks with statistical channel state information,” \emph{IEEE Trans. Sig. Proc.}, vol. 66, no. 22, pp. 5902--5916, Nov. 15, 2018.

\bibitem{song12}
C. Song, K.-J. Lee, and I. Lee, ``MMSE-based MIMO cooperative relaying systems: Closed-form designs and outage behavior,'' \emph{IEEE J. Sel. Areas Commun.}, vol. 30, no. 8, pp. 1390–1401, Sep. 2012.

\bibitem{kong14}
H.-B. Kong, C. Song, H. Park, I. Lee, ``A New Beamforming Design for MIMO AF Relaying Systems With Direct Link,'' \emph{IEEE Trans. Commun.}, vol. 62, no. 7, pp. 2286--2295, July 2014.

\bibitem{he16}
Z. He, J. Zhang, W. Liu, and Y. Rong, ``New results on transceiver design for two-hop amplify-and-forward MIMO relay systems with direct link,'' \emph{IEEE Trans. Signal Processing}, vol. 64, no. 20, pp. 5232--5241, Oct. 2016.

\bibitem{horn}
R.\ Horn and C.\ Johnson, {\em Matrix Analysis (2nd ed.)}. New York: Cambridge University Press, 2013.


\bibitem{lars}
E.G.\ Larsson and P.\ Stoica, \emph{Space-time Block Coding for Wireless Communications}. Cambridge University Press, 2003.

\bibitem{boyd}
S.\ Boyd and L.\ Vandenberghe, {\em Convex Optimization}. Cambridge University Press, 2004.

\bibitem{cover}
T.M.\ Cover and J.A.\ Thomas, \emph{Elements of Information Theory}. New York: Wiley, 2006.

\bibitem{fied71}
M.\ Fiedler, ``Bounds for the determinant of the sum of Hermitian matrices,'' \emph{Proc.\ Amer.\ Math.\ Soc.}, vol.~30, no.~1, Sept.~1971, pp.~27--31.

\bibitem{princeton}
T.\ Gowers, \emph{The Princeton Companion to Mathematics}. Princeton University Press, 2008.

\bibitem{tel99}
I.E.Telatar, ``Capacity of multi-antenna Gaussian channels,'' \emph{European Trans. Telecommun.}, vol.10, no.6, pp.585--595, Nov.1999.

\bibitem{isit21}
G.\ Taricco, ``Information Rate Optimization for Joint Relay and Link in Non-Regenerative MIMO Channels,'' \emph{IEEE ISIT 2021}.



\end{thebibliography}
\end{document}